# Local-gated single-walled carbon nanotube field effect transistors assembled by AC dielectrophoresis

**Paul Stokes and Saiful I. Khondaker***

Nanoscience Technology Center & Department of Physics, University of Central Florida, 12424 Research Parkway, Orlando FL 32826, USA

* To whom correspondence should be addressed. E-mail: saiful@mail.ucf.edu

**Abstract**

We present a simple and scalable technique for the fabrication of solution processed & local gated carbon nanotube field effect transistors (CNT-FETs). The approach is based on directed assembly of individual single wall carbon nanotube from dichloroethane via AC dielectrophoresis (DEP) onto pre-patterned source and drain electrodes with a local aluminum gate in the middle. Local-gated CNT-FET devices display superior performance compared to global back gate with on-off ratios $>10^4$ and maximum subthreshold swings of 170 mV/dec. The local bottom-gated DEP assembled CNT-FETs will facilitate large scale fabrication of complementary metal-oxide-semiconductor (CMOS) compatible nanoelectronic devices.

## 1. Introduction

Carbon nanotube field effect transistors (CNT-FETs) have displayed exceptional electrical properties that are superior to the traditional silicon metal-oxide-semiconductor field effect transistor (MOSFET) without the problem of scaling down [1,2]. Early fabrication techniques of CNT-FETs involved random placements of CNT either on pre-patterned electrodes or by dispersing them on substrates, locating them with atomic force microscopy (AFM), and finally defining source and drain electrodes using electron beam lithography (EBL) [3]. In addition, most of these devices were often controlled by a global back gate because of its ease of processing. Such fabrication processes neither offer high throughput nor individual control of each CNT-FET necessary for parallel fabrication of nanoelectronic devices.

For large-scale fabrication of CNT-FET devices three conditions need to be satisfied: (i) separation of semiconducting and metallic carbon nanotubes must be realized, (ii) nanotubes need to be assembled at selected positions of the circuit with high yield, and (iii) each nanotube must be addressed individually with a local gate. While control over separation remains elusive, significant progress has been made in directed assembly of CNT-FET by patterning catalyst for the chemical vapor deposition (CVD) growth process and using a local top gate [4-7]. Although devices made from such techniques show the best performance so far, however, CVD requires the growth temperature to be $900^0$ C which is prohibitively high for current CMOS fabrication technologies. Other assembly techniques such as chemical and biological patterning [8, 9], flow assisted alignment [10], Langmuir-Blodgett assembly [11], bubble blown films [12], and contact printing [13] demonstrated for 1D nanostructures may also provide route for large scale fabrication of CNT-FET.

Recently, AC dielectrophoresis (DEP) has been utilized for large scale assembly of individual single-walled carbon nanotube (SWNT) or bundles at selected positions of the circuit [14-25]. In DEP, CNTs are assembled from solution using a non-uniform AC electric field. However, all DEP assembled CNT-FETs reported in the literature used only a global back gate [14-19,21-22, 24]. Global back gated devices give poor device performance due to inefficient gate coupling and contact-controlled operation. This means when the back gate is active, it controls the Schottky barriers rather than the conducting channel itself, causing slow-switching behavior [26]. In addition, a global back gate cannot address CNT-FETs individually, making integrated circuits out of the question.

Here we report on the fabrication and device characteristics of local bottom-gated DEP assembled CNT-FETs. First, gold (Au) source and drain electrodes of 1 μm separation with a 100 nm wide aluminum (Al) gate electrode in the middle were fabricated with standard optical and electron beam lithography (EBL). Carbon nanotubes suspended in dichloroethane (DCE) were then assembled between source and drain electrodes via DEP. We find that both metallic and semiconducting nanotubes can be assembled and the centered aluminum gate does not affect the DEP assembly. We also show that the measured device performance such as subthreshold swing of local-gated semiconducting nanotube FET is superior compared to the global back gated device possibly due to channel controlled operation. Local bottom gated DEP assembled CNT-FET will facilitate large scale fabrication of CNT based integrated circuits and other nanoelectronic devices such as sensors.

## 2. Experimental Details:

Devices were fabricated on heavily doped silicon (Si) substrates capped with a thermally grown 250 nm thick $SiO_2$ layer. Figure 1 presents an illustration of the device fabrication procedure. The electrode patterns were fabricated by a combination of optical and electron beam lithography (EBL). First, contact pads and electron beam markers were fabricated with optical lithography using double layer resists (LOR 3A/Shipley 1813) developing in CD26, followed by thermal evaporation of chromium (Cr) (5 nm) and Au (50 nm) and finally standard lift-off. Source and drain electrode patterns were defined with EBL using single layer PMMA resists and then developing in (1:3) methyl isobutyl ketone:isopropal alchohol (MIBK:IPA). After defining the patterns, 5 nm Cr and 20 nm thick Au were thermally deposited followed by lift-off. The source and drain electrodes are chosen to be tapered shape as shown in figure 1a to maximize the electric field at the sharp edge and increase the chance of obtaining an individual SWNT connection during the DEP assembly. EBL was implemented once again to define the Al gate patterns using single layer PMMA resist and developing in MIBK:IPA following thermal deposition of 20-25 nm of Al and lift-off (figure 1b). Al gate patterns had a partial overlap with selected gold contact pads defined earlier in order to apply a voltage to the local gate. The sample is finally treated in oxygen plasma for 10 minutes to ensure a good 2-3 nm thick aluminum oxide ($Al_2O_3$) layer. The DEP assembly of CNT shown in Fig 1c was carried out as follows:

A very small amount (~0.3 μg) of highly purified HiPco grown SWNTs (purchased as nanotube soot from Carbon Nanotechnologies Inc.) was ultrasonically dispersed in 5 ml of 1,2-dichloroethane for approximately 30 minutes. Immediately after the dispersion is complete, a small drop (~8ul) was cast onto a chip with 12 pairs of source-drain electrodes, each containing a 100 nm wide Al gate. An AC voltage of approximately 8 $V_{P-P}$ at 1 MHz is applied with a function generator for 1-2 seconds to the electrode pair and then moved to the next pair in a probe station. It has been shown that when metallic posts are present in between source and drain electrodes they may influence the DEP assembly process for CNT [18]. We have found that our gate electrode does not influence the ability to assemble CNTs between the 1 μm gap possibly due to the Al$_2$O$_3$ insulating layer. We have studied other AC voltages and trapping times and find that larger voltages applied in shorted time periods give us more control over the ability to assemble an individual carbon nanotube or a small diameter bundle. The AC voltage gives rise to a time averaged dielectrophoretic force. For an elongated object it is given by

$$F_{DEP} \propto \varepsilon_m \text{Re}[K_f]\nabla E^2_{RMS}, \quad K_f = \frac{\varepsilon^*_p - \varepsilon^*_m}{\varepsilon^*_m}, \quad \varepsilon^*_{p,m} = \varepsilon_{p,m} - i\frac{\sigma_{p,m}}{\omega},$$

where $\varepsilon_p$ and $\varepsilon_m$ are the permittivity of the nanotube and solvent respectively, $K_f$ is the Claussius-Mossotti factor, $\sigma$ is the conductivity, and $\omega = 2\pi f$ is the frequency of the applied AC voltage [27]. The induced dipole moment of the nanotube interacting with the strong electric field causes the nanotubes to move in a translational motion along the electric field gradient [23,24]. Because of the strong dielectrophoretic force, the nanotubes are reproducibly aligned at the tips of the source and drain electrodes where the electric field is maximum. This is shown in figure 1d where we present a representative AFM image of one of our devices. After the trapping process is complete, the chip is rinsed with IPA and blow dried with nitrogen gas to remove any unwanted nanotubes or impurities in the suspension. Other groups have dispersed CNT in an aqueous sodium dodecyl sulfate (SDS) [14-20, 23], isopropyl alcohol (IPA) [21], dimethylformamide (DMF) [24], and DCE [25] for DEP assembly. Although an aqueous SDS solution has been used most frequently and has been shown to disperse CNT well, it is not compatible with our device fabrication process. The aluminum oxide is known to interact with water and the gate dielectric will be damaged if an aqueous SDS solution is used. We choose DCE because it is a common solvent used to disperse CNTs and does not interact with the Al$_2$O$_3$ gate dielectric.

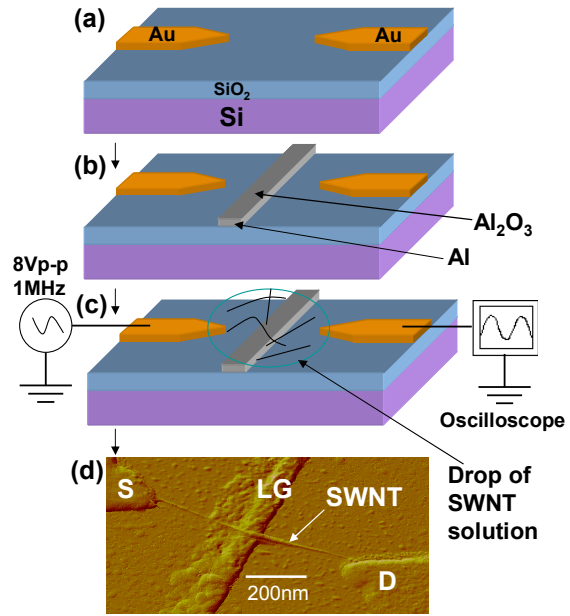

**Figure 1.** Fabrication of local gated CNT-FET device. (a) Source (S) - drain (D) electrodes of 1 μm separation are patterned on heavily doped Si/$SiO_2$ substrates (250 nm thick oxide layer). (b) Local Al gate electrodes are patterned using EBL and a 2-3 nm thick $Al_2O_3$ is created by oxygen plasma treatment. (c) DEP assembly of CNT. An AC voltage of 8$V_{p-p}$ is applied for 1-2 seconds to the source electrode with a function generator (d) Resulting AFM image of a device showing nanotubes are assembled at the tips.

### 3. Results and Discussion

After the DEP assembly, electronic transport measurements were carried out in a probe station and ambient environment. Out of 115 devices that we have tested, we find that in ~35% cases the electrodes were bridged with either an individual SWNT or a small diameter CNT bundle determined by AFM measurements. The rest of the samples either have multiple connections or no connection at all. Here we focus on devices containing individual nanotubes or small bundles. The two terminal resistance of our samples is usually in the range of 1-10 Mohm. Our value of contact resistance is consistent with other DEP assembled devices [14-25], however, higher than that of top contact CVD grown devices (~ 100 kOhm), where the nanotube is first grown by CVD on a substrate, made contact to, and then annealed. The contact resistance depends on several factors such as work function of the metal being used, diameter of the nanotubes, surface properties, contact area and device geometry. From our observation of many AFM images, the DEP assembled devices seem to be end-contacted (the nanotube's end is connected to the electrode's end). Therefore the contact area is very small. In addition, we used gold as a metal electrode which gives a larger Schottky barrier at metal – CNT interface. These may be a few reason for large contact resistance in our DEP assembled devices. We are currently investigating ways of reducing contact resistance such as using palladium contact, post deposition and annealing.

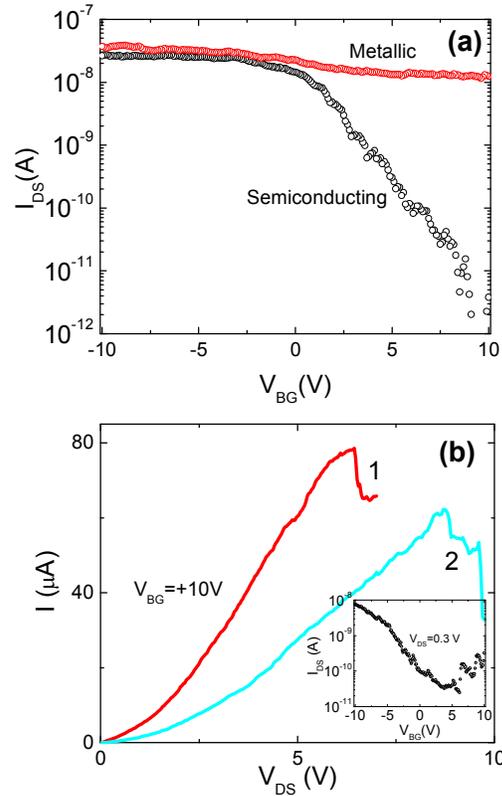

**Figure 2.** a) Drain current ($I_{DS}$) - back gate voltage ($V_{BG}$) characteristics of representative DEP assembled metallic and semiconducting devices ($V_{DS}$ = 0.3 V). (b) Transformation of a metallic nanotube bundle into a semiconducting device by selective burning by sequential ramps of $V_{DS}$ (labeled 1 and 2) with back gate set to 10V. INSET: Resultant current - back gate characteristic after burning ($V_{DS}$ = 0.3 V).

Before discussing local-gated devices, we first present electronic characteristics of global back gated devices. We used heavily doped Si substrates as a global back gate. Figure 2a shows drain current ($I_{DS}$) as a function of back gate voltage ($V_{BG}$) of representative metallic and semiconducting devices for a fixed source drain voltage ($V_{DS}$) of 0.3 V. The metallic single wall nanotubes (m-SWNTs) show weak modulation in $I_{DS}$ as a function of $V_{BG}$, whereas semiconducting single wall nanotubes (s-SWNTs) show several orders of magnitude change in $I_{DS}$ as a function of $V_{BG}$. About half of the devices we examined for this study display p-type semi-conducting behavior characterized by $V_{BG}$ following the DEP assembly. These devices displayed current on-off ratios >$10^4$ and subthreshold swings $S = [dV_G / d(\log(I_{DS}))]$ from 1000-2500 mV/dec. Other DEP assembled CNT-FET devices reported in the literature also display similar device characteristics [14-15, 17-19, 21-22] The other half of the devices we examined showed metallic behavior where ~30% of them were able to be transformed to semi-conducting via selective burning of m-SWNTs contained in the bundles [28]. To deplete carriers in the s-SWNTs, 10 V is applied to the $V_{BG}$ and then $V_{DS}$ is ramped up to approximately 10 V until the current starts to drop as shown in figure 2b labeled 1 and 2. Some devices may be destroyed if further burning is done (usually the current will eventually drop to zero at voltages greater than 15 V). The $I_{DS}$ - $V_{BG}$ characteristics are measured once again after each burning step and the procedure is repeated until a larger on-off ratio is observed. The final back-gate voltage dependence is shown in figure 2b's inset with an on-off ratio of ~1000. Most of the devices that needed selective burning display lower on-off ratios compared to the as assembled semiconducting CNT-FETs. The fair performance can be attributed to the presence of metallic nanotubes still within the bundles between the electrodes introducing more scattering into the transport.

Figure 3a presents characteristics of one of our local-gated device where we also present back-gated data of the same device for comparison. It can be clearly seen that the threshold voltage for back gate is 10 V while for local gate it is only 1 V, indicating an extremely better gate coupling for local gate. Additionally, the back gate has a broad maximum subthreshold swing of ~ 2230 mV/dec, whereas the local gate has a value of ~ 170 mV/dec, demonstrating much faster switching behavior by the local gate. Figure 3a's inset displays an expanded view of the local gate dependence, clearly showing up to ~ 4 orders of magnitude change in $I_{DS}$ for a small gate voltage range. The leakage current measured for our device is negligible (<1 pA for a voltage of -2V to +2V applied to the local gate). Other local-gated devices that we have fabricated show similar FET response. Small values of the subthreshold swing and low threshold voltage are preferred in FETs for low power consumption and high speed operation [5,29]. In figure 3b we plot $I_{DS}$ versus $V_{DS}$ up to the saturation regime at different local gate voltages (0 to -1.4 V in steps of 0.2 V from top to bottom). These output characteristics are similar to typical p-MOSFET devices. From here we can calculate the transconductance in the saturation regime by taking $g_m = (dI_{DS}/dV_g)_{V_{DS}=-1.6V}$ to be 0.3 μS. Normalized by the width of the nanotube (~ 1.5 nm), we derive the normalized transconductance for the local gate of 200 S/m, while for the back gate we derive a value of 3.3 S/m. This also indicates better efficiency of the local aluminum gate. These local gate characteristics are superior to other DEP assembled CNT-FETs [16] reported in the literature and comparable to some higher performance CNT-FET devices reported recently with $Al_2O_3$ gates [6,29] and high-K dielectrics [5]. Table 1 gives a brief comparison of a few recently fabricated CNT-FETs along with the device presented here.

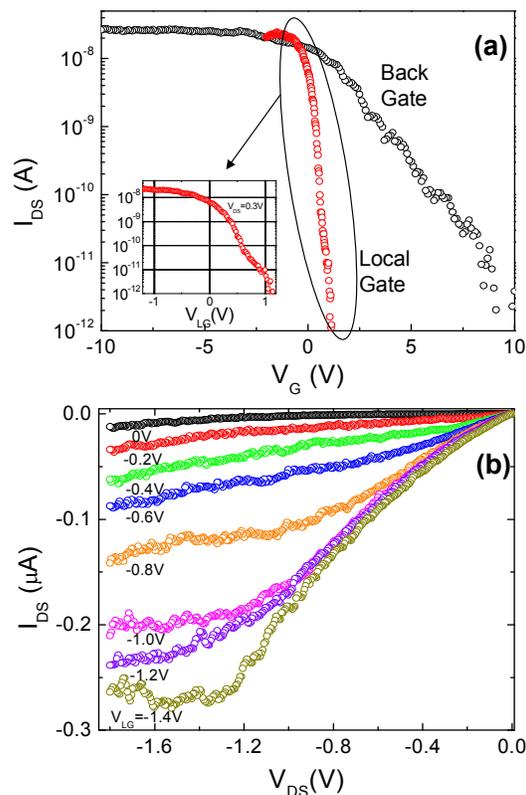

**Figure 3.** (a) Drain current versus local gate voltage and back gate voltage for comparison form the same device after DEP assembly. $V_{DS}$ = 0.3 V for both curves. Local gate shows far better gate coupling. INSET: Expanded plot of $I_{DS}$ vs. $V_{LG}$ showing low threshold voltage and subthreshold swing. The gate leakage current is < 1 pA. b) Output characteristics, $I_{DS}$ vs. $V_{DS}$ for different gate voltages up to the saturation regime.

| Assembly Method | Gate Material | Subthreshold Swing (mV/dec) | Threshold Voltage (V) | Ref |
|---|---|---|---|---|
| DEP | $SiO_2$/Back gate | 1200 | 1.5 | 16 |
| CVD/ALD | $Al_2O_3$/Local gate | 120 | N/A | 29 |
| CVD | $SiO_2$/Back gate | 1000-2000 | N/A | 5 |
| DEP | $Al_2O_3$/Local gate | 170 | 1 | this work |

**Table 1.** Comparison of a few recent CNT-FETs assembled by DEP and CVD technique.

A possible reason for the fast switching behavior of our local gated device may be due to channel controlled operation [26]. There are two accepted sources of operation for CNT-FETs, (i) contact-controlled operation from formation of Schottky barriers at the contacts [30] or (ii) channel controlled operation [31] (typically for good Ohmic contacted devices). For our local-gated device, the mechanism is channel controlled owed to the thin Al gate in the middle which is relatively far away from the source and drain electrodes. When the Al gate is active its electric field can not affect the contact between the nanotube electrodes. Thus, the switching will be due to the local gate controlling the channel and not depend on whether the contact has a large Schottky barrier or if it is an ohmic contact. We are currently working on scaling down the gate length further to increase device performance.

**4. Conclusions**
In conclusion, we have fabricated CNT-FETs with local Al bottom gates through DEP. Our method offers a convenient way to assemble local-gated CNT-FET devices from solution without the need of high temperature growth, making it compatible with present microfabrication technology. Our local-gated devices show superior characteristics such as small values of threshold swing and low threshold voltage compared to other DEP assembled back gated CNT-FETs. Local gating offers fast switching behavior due to the channel controlled mechanism owed to the thin local Al gate. Directed assembly of local gated CNT-FETs at selected position of the circuit via DEP pave the way for large scale fabrication of CMOS compatible nanoelectronic devices.

**References**

1. McEuen P L and Park J 2004 *MRS Bulletin* **29** 272
2. Avouris P 2007 *Physics World* **20** 40
3. Tans S J, Verschueren A R M and Dekker C 1998 *Nature* **343** 49
4. Javey A, Wang Q, Ural A, Li Y and Dai H 2002 *Nano Lett.* **2** 929
5. Javey A, Kim H, Brink M, Wang Q, Ural A, Guo J, McIntyre P, McEuen P, Lundstrom P and Dai H 2002 *Nature Mater.* **1** 241
6. Kim S K, Xuan Y, Ye P D, Mohammadia S, Back J H and Shim M 2007 *Appl. Phys. Lett.* **90** 163108
7. Kang S J, Kocabas C, Ozel T, Shim M N, Pimparkar M, Alam A, Rotkin S V and Rogers J A 2007 *Nature Nanotech.* **2** 230
8. Auvray S, Derycke V, Goffman M, Filoramo A, Jost O, and Bourgoin J-P 2005 *Nano Lett.* **5** 451
9. Keren, K, Berman R S, Buchstab E, Sivan U and Braun, E 2003 *Science* **302** 1380
10. Huang Y, Duan X, Wei Q, Lieber C M 2001 *Science* **291**, 630
11. Jin S, Whang D, McAlpine M C, Friedman R S, Wu Y, and Lieber C M 2004 *Nano Lett.* **4** 915
12. Guihua Y, Anyuan C and Lieber C M 2007 *Nature Nanotech.* **2** 372
13. Javey A, Nam S W, Friedman R S, Yan H and Lieber C M 2007 *Nano Lett.* **7**, 773
14. Vijayaraghavan A, Blatt S, Weissenberger D, Oron-Carl M, Hennrich F, Gerthsen D, Hahn H and Krupke R 2007 *Nano Lett.* **7** 1556
15. Dong L F, Youkey S, Bush J and Jiao J 2007 *J. Appl. Phys* **101** 024320
16. Dong L F, Chirayos V, Bush J, Jiao J, Dubin V M, Chebian R V, Ono Y, Conley Jr J F and Ulrich B D 2005 *J. Phys. Chem. B* **109** 13148.
17. Zhang Z, Liu X, Campbell E E B and Zhang S 2005 *J. Appl. Phys.* **98** 056103
18. Banerjee S, White B, Huang L, Rego B J, O'brien S and Herman I P 2007 *Appl. Phys. A* **86** 415
19. Banerjee S, White B, Huang L, Rego B J, O'brien S and Herman I P 2006 *J. Vac. Sci. Tech. B* **24** 3173
20. Krupke R, Hennrich F, Kappes M M and Lo1hneysen H v 2004 *Nano Lett.* **4** 1395
21. Li J, Zhang Q, Yang D and Tian J 2004 *Carbon* **42** 2263
22. Zhanga Z, Zhang S and Campbell E E B 2006 *J. Vac. Sci. Technol. B* **24** 131
23. Krupke R, Hennrich F, Lohneysen H v, Kappes M M 2003 *Science* **301** 344
24. Krupke R, Hennrich F, Weber H B, Kappes M M and Lohneysen H v 2003 *Nano Lett.* **3** 1019
25. Seo H-W, Han C-S, Choi D-G, Kim K-S and Lee Y-H 2005 *Microelec. Eng.* **81** 83
26. Park J-Y 2007 *Nanotech.* **18** 095202
27. Jones, T. B. *Electromechanics of Particles*; Cambridge University Press: Cambridge, 1995.
28. Collins P G, Arnold M S and Avouris Ph 2001 *Science* **292** 706
29. Lin Y M, Appenzeller J, Chen Z, Chen Z-G, Cheng H-M and Avouris Ph 2005 *IEEE Elec. Dev. Lett.* **26** 823
30. Heinze S, Tersoff J, Martel R, Derycke V, Appenzeller J and Avouris Ph 2002 *Phys. Rev. Lett.* **89** 106801
31. Javey A, Guo J, Wang Q, Lundstrom M and Dai H 2003 *Nature* **424** 654